\documentclass[%
 prl,
 jmp,%
 amsmath,amssymb,
 reprint,%
]{revtex4-1}

\usepackage{graphicx}
\usepackage{dcolumn}
\usepackage{bm}
\usepackage{amsmath}

\begin{document}






\title{Liquid-state acoustically-nonlinear nanoplasmonic source of optical frequency combs}
\author{Ivan S. Maksymov}
\author{Andrew D. Greentree}
\affiliation{ARC Centre of Excellence for Nanoscale BioPhotonics, School of Science, RMIT University, Melbourne, VIC 3001, Australia}

\date{\today}

\begin{abstract}
Nonlinear acoustic interactions in liquids are effectively stronger than nonlinear optical interactions in solids. Thus, harnessing these interactions will offer new possibilities in the design of ultra-compact nonlinear photonic devices. We theoretically demonstrate a hybrid, liquid-state and nanoplasmonic, source of optical frequency combs compatible with fibre-optic technology. This source relies on a nanoantenna to harness the strength of nonlinear acoustic effects and synthesise optical spectra from ultrasound. 
\end{abstract}

\maketitle

An optical frequency comb (OFC) is a spectrum consisting of a series of discrete, equally spaced elements \cite{Tor14, Sav16}. OFCs are important in precision measurements, microwave generation, telecommunications, astronomy, spectroscopy and imaging \cite{Tor14}. They are usually produced by using mode-locked lasers or exploiting nonlinear (NL) optical effects in optical fibres \cite{Tor14}. On-chip integrated NL optics \cite{Kip11, Lu16, Sav16} and optomechanics \cite{Sav11} also allow generating OFCs in microphotonic systems. However, such sources have a number of limitations that motivate the ongoing research efforts \cite{Tor14, Sav16}.    

Nanoscale OFC sources are even more difficult to achieve. The development of such sources faces the same challenges as in microphotonic systems \cite{Tor14, Sav16}. Moreover, whereas NL optical effects may in general be enhanced using ultra-small metal nanostructures \cite{Kau12}, the generation of OFCs at the nanoscale is impeded by high absorption, short NL interaction lengths, phase matching challenges, and other fundamental limitations.

These roadblocks can be mitigated by capitalising, for example, on the advances in NL diamond photonics \cite{Hau14} and plasmonic nanoantennae (NAs) \cite{Mak12, Xio16}. An OFC may be generated with a tapered multielement NA designed such that its elements resonate at one of the comb frequencies \cite{Mak12}. However, tapered NAs have a relatively large metal volume, which decreases the NL generation efficiency because of absorption and heating. On the other hand, diamond combines a high refractive index, low absorption losses and excellent thermal properties attractive for integrated NL photonics \cite{Hau14}. However, its optical nonlinearities have not yet been fully studied, but the footprint of diamond-based devices remains in the range of several hundreds of square microns \cite{Hau14}. Thus, the search for new nanoscale OFC sources remains open.

Here, we theoretically demonstrate an alternative scheme for synthesis of OFCs with ultra-small, single-element NAs. We turn our attention to the fact that NL acoustic effects in liquids \cite{Gurbatov} are much stronger than their optical counterparts. The propagation of the incident ultrasound with frequency $f_{\rm{us}}$ in an acoustically NL medium gives rise to higher harmonics with frequencies $nf_{\rm{us}}$ ($n$ is a positive integer). In contrast to optics where the generation of the higher ($n>5$) harmonics is often difficult to observe, acoustic harmonics with $n$ up to $\sim 15$ are readily achievable. Many liquids have a large NL coefficient $\beta$, which is the key parameter to characterise the strength of the NL acoustic effects. For example, water has $\beta = 3.5$, optically transparent oils have $\beta \approx 6$, but water with bubbles has $\beta \approx 5000$ \cite{Gurbatov}.

At present there are a number of technological challenges that prevent the practical integration of liquid-state elements into traditional solid-state microphotonic systems \cite{Eun12, Phi14}. However, liquid-state devices offer a number of potentially transformative advantages for microphotonic systems \cite{Eun12, Phi14}. Successful examples include liquid-state optical lenses \cite{Kru03}, dye lasers \cite{Li06}, and liquid-core optical fibres \cite{Alt97}.

\begin{figure}[t]
\centering\includegraphics[width=7.5cm]{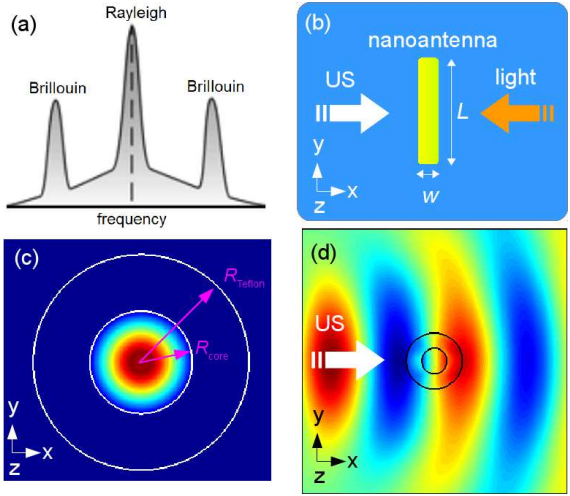}
\caption{ (a) Sketch of a typical BLS spectrum. (b) Schematic of a nanorod NA immersed into water and insonated by ultrasound (US). The wavelength of ultrasound is two orders of magnitude larger than the width $w=30$~nm and length $L=340$~nm of the NA. The E-field of the optical plane wave incident on the NA is polarised along the $y$-coordinate. (c) $|E|$-field intensity in the cross-section of the water-filled core ($R_{\rm{core}} = 250$~$\mu$m) fibre with a Teflon coating ($R_{\rm{Teflon}} = 525$~$\mu$m). The fibre is surrounded by water. The optical frequency is $405$~THz. (d) Instantaneous time-domain snapshot of the $T_{\rm{xx}}$ stress component in the cross-section of the fibre and surrounding water. Note that the incident ultrasound wavefront is almost unchanged by the fibre. The optical and elastic finite-difference time-domain (FDTD) methods were used to obtain the results in (c, d).}
\label{fig:fig0}
\end{figure}

In liquids, ultrasound may optically be detected through the Brillouin Light Scattering (BLS) effect \cite{Fab68}. The spectrum of light scattered from single-frequency incident ultrasound with frequency $f_{\rm{us}}$ has a form of a triplet [Fig.~\ref{fig:fig0}(a)], consisting of the central Rayleigh peak and two BLS peaks shifted by $\pm f_{\rm{us}}$. It has been demonstrated that the sensitivity of light to single-frequency ultrasound may be increased by using plasmonic nanostructures including NAs \cite{Utegulov_new, Mak16}. 

In the presence of both incident ultrasound and its NL-generated harmonics, the BLS spectrum exhibits multiple, equally spaced peaks at $\pm nf_{\rm{us}}$. Moreover, the basic criteria for the efficient OFC generation, such as high spectral coherence and the possibility to synthesise the spacing between the spectral lines \cite{Tor14}, are in general met in NL acoustics and used, e.g., to measure $\beta$ of NL acoustic media \cite{Duq97}. However, despite a large strength of the NL acoustic effects, the intensity of the BLS peaks is intrinsically low \cite{Fab68} and decreases as $n$ is increased. Consequently, to be employed as an OFC, the intensity of each peaks in the BLS spectrum has to be increased.

In this Letter, we demonstrate that the intensity of all peaks in the BLS spectrum of NL ultrasound may be increased by using an ultrasmall, single-element plasmonic NA. We show that an increase in the BLS peak intensity allows synthesising an OFC with the spacing between the spectral lines controllable at will by changing the frequency of ultrasound. 

We consider a single, square cross-section silver nanorod NA with an optimised \cite{Mak16} $30 \times 340 \times 30$~nm$^3$ volume [Fig.~\ref{fig:fig0}(b)]. The NA is surrounded by water with refractive index $n_{\rm{water}}=1.33$, NL acoustic parameter $\beta_{\rm{water}} = 3.5$, density $\rho = 1000$~kg/m$^3$ and speed of sound $c_{\rm{0}} = 1500$~m/s \cite{Mak16}. Because the majority of optically transparent liquids have $\beta \approx \beta_{\rm{water}}$ \cite{Gurbatov}, results presented below will hold for other liquids including oils.

Since the integration of liquid-state and solid-state microphotonics is a challenge, we discuss a strategy for the integration of this NA with optical fibres. We propose to use a liquid core optical fibre \cite{Alt97}, which may host the NA inside the core. Because simultaneous plasmonic-opto-acoustic simulations of this scenario are computationally impractical, we first model optical and acoustic properties of the fibre without the NA. By considering a Teflon water-core fibre ($n_{\rm{Teflon}}=1.29$ \cite{Alt97}, $\rho = 2200$~kg/m$^3$, $c_{\rm{0}} = 1400$~m/s and shear wave speed $440$~m/s) we show that the fibre may operate in water [Fig.~\ref{fig:fig0}(c)]. We also show that a good agreement between the characteristic specific acoustic impedance ($z_{\rm{0}} = \rho c_{\rm{0}}$) of Teflon and water makes this fibre mostly transparent to ultrasound propagating in the direction perpendicular to the fibre core [Fig.~\ref{fig:fig0}(d)].

\begin{figure}[t]
\centering\includegraphics[width=8.5cm]{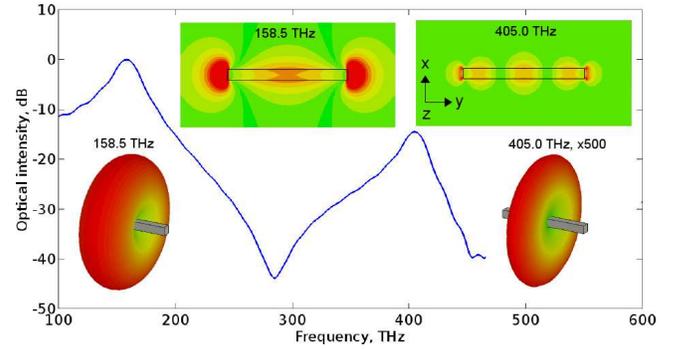}
\caption{ Optical properties of the NA. The insets show the near-field E-field profiles (top) and far-field power profile (bottom) corresponding to the fundamental ($158.5$~THz) and higher-order ($405$~THz) modes. The far-field power profile of the higher-order mode is multiplied by $500$.}
\label{fig:fig1}
\end{figure}

$3$D simulations with CST Microwave Studio software reveal that in the $100-500$~THz spectral range the spectrum of the NA [Fig.~\ref{fig:fig1}(a)] has the fundamental mode and one higher-order mode. In \cite{Mak16} we established that the results of $3$D simulations are qualitatively reproduced by $2$D simulations, with the major difference being a blueshift of all resonance peaks with respect to the $3$D case, because the nanorod is modelled as a plate. 

The NA tuned on the higher-order mode radiates lower power into the far-field zone (see the bottom insets in Fig.~\ref{fig:fig1} and note the $\times 500$ zoom at $405$~THz) as compared with the fundamental mode. This leads to a stronger near-field confinement to the metal surface of the NA as compared with the fundamental mode (top insets).

Because the dielectric permittivity of water and silver are modulated by ultrasound as $\Delta \epsilon(t) \propto \nabla \cdot \bm{s}(t)$ with $\bm{s}(t)$ the time-dependent displacement field of ultrasound \cite{Mak16}, a stronger near-field confinement leads to an increased sensitivity to ultrasound as compared with the case of bulk water without the NA. In bulk water, light may sense ultrasound only when the two waves co-propagate over macroscopically long distances \cite{Fab68}. However, with the NA tuned at its higher order mode this condition is lifted \cite{Mak16}.

We solve the Earnshaw equation to analyse the weak NL acoustic interaction in a liquid \cite{Gurbatov}.

\begingroup\makeatletter\def\f@size{9}\check@mathfonts
\def\maketag@@@#1{\hbox{\m@th\large\normalfont#1}}%
\begin{align}
\frac{\partial^{2} \xi}{\partial t^{2}} = c^{2}_{\rm{0}} \frac{\partial^{2} \xi / \partial x^{2}}{(1+\partial \xi / \partial x)^{\gamma+1}}
\label{eq:one}.
\end{align}\endgroup

\noindent where $c_{\rm{0}}$ is the speed of sound in the liquid and $|\partial \xi / \partial x| << 1$. We expand the term $(1+\partial \xi / \partial x)^{-(\gamma+1)}$ into the binomial series and employ the method of slowly varying envelope to find a solution in the form of two arbitrary travelling waves $\xi = \Phi(t-x/c_{\rm{0}}) + \Psi(t+x/c_{\rm{0}})$. We consider only the wave travelling in the positive direction. The wave profile changes slowly leading to $\xi = \Phi(\tau = t-x/c_{\rm{0}}, x_{\rm{1}}=\mu x$), where $\mu<<1$. 

Following \cite{Gurbatov} we obtain the simple wave equation $
\frac{\partial u}{\partial x} = \frac{\beta}{c^{2}_{\rm{0}}} u \frac{\partial u}{\partial \tau}$, where $u = \partial \xi / \partial \tau$ is the particle velocity and $\beta = \frac{\gamma+1}{2}$. In the following, $\beta$ defines the normalised dimensionless distance $z=(\beta/c^{2}_{\rm{0}}) \omega u_{\rm{0}} x = x/x_{\rm{S}}$, where $\omega = 2 \pi f_{\rm{us}}$ and $x_{\rm{S}}$ is the acoustic wave discontinuity formation length \cite{Gurbatov}.

\begin{figure}[t]
\centering\includegraphics[width=8.5cm]{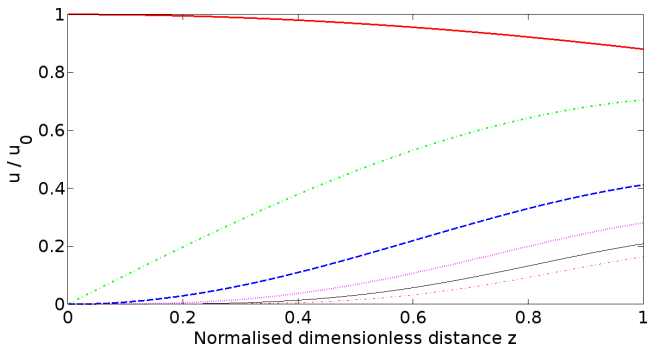}
\caption{ Distance-dependent amplitudes of the first six, from top to bottom, harmonics of an initially sinusoidal ultrasound wave. The second and higher harmonic amplitudes are multiplied by $2$ for the sake of visualisation.}
\label{fig:fig2}
\end{figure} 

We use the Fourier series expansion to derive the relationship between the harmonic amplitude $u$ and the amplitude of the incident ultrasound wave $u_{\rm{0}}$ as a function of $z$. Then we calculate the expansion coefficients by using the $n$th-order Bessel functions of the first kind to arrive to the Bessel-Fubini solution 

\begingroup\makeatletter\def\f@size{9}\check@mathfonts
\def\maketag@@@#1{\hbox{\m@th\large\normalfont#1}}%
\begin{align}
\frac{u}{u_{\rm{0}}} = \sum_{n=1}^{\infty} \frac{2J_{\rm{n}}(nz)}{nz} \sin(n \omega \tau)
\label{eq:five}.
\end{align}\endgroup

Figure~\ref{fig:fig2} shows the dependencies of the first six harmonic amplitudes of ultrasound on the distance $z$. At $z=1$ these waves carry $\sim 98\%$ of the total acoustic energy, with the remainder being carried by the higher-order harmonics.

\begin{figure*}[t]
\centering\includegraphics[width=12cm]{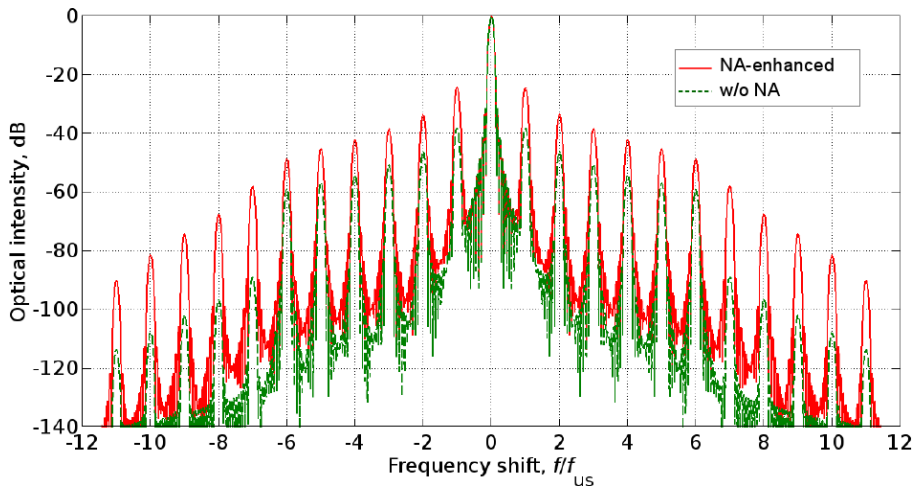}
\caption{ OFC synthesised from NL ultrasound by the BLS effect. The optical optical intensity is shown as a function of the frequency shift with respect to $405$~THz (the frequency of incident light), normalised to the frequency $f_{\rm{us}}$ of the incident ultrasound. Red solid line: NA-enhanced spectrum. Green dashed line: without the NA.}
\label{fig:fig3}
\end{figure*}

The result presented in Fig.~\ref{fig:fig2} is valid for a broad range of frequencies $f_{\rm{us}}$ \cite{Fab68}. We choose $f_{\rm{us}}$ such that the frequency of the $10$th harmonic remains below $\sim 10-20$~GHz, which is the frequency above which the NA may resonantly vibrate \cite{Sav09}. Such vibrations may lead to interesting effects, including the generation of OFCs \cite{Kip11, Lu16}. However, we will mostly be interested in the operation of the NA below its structural resonances -- in the quasi-static regime. Thus, in our analysis the maximum value of $f_{\rm{us}}$ will be $1-2$~GHz. In liquids at the pressure levels up to several GPa, the speed and absorption coefficient of GHz-range acoustic waves are similar to those of ultrasound \cite{Fab68}. Such waves can be generated by modern piezoelectric transducers. 

In the quasi-static regime the NL acoustic effects are unaffected by the NA \cite{Mak16}. The NA is considered as a bulk medium, the normal stress components are equal to the ultrasound pressure $P$ taken with the opposite sign, and the shear stress is zero. From Hooke's law we define the strain components, which are used to calculate the displacement $\bm{s}$ and the respective variations $\Delta \epsilon$ of the dielectric permittivity of water and silver due to the ultrasound pressure \cite{Mak16}. 

To generate an OFC we have to employ high-intensity ultrasound to produce as high as possible acoustic intensities of NL harmonics. However, for the presented NL acoustic theory to be exact we have to maintain $\mu<<1$ (i.e. the acoustic Mach number $M=u_{\rm{0}}/c_{\rm{0}}<<1$), which restricts the maximum value of the pressure that we can choose. Thus, as a trade-off, in the following we assume that $P=500$~MPa, which corresponds to $M \approx 0.2$ for which we expect our model to remain mostly valid.

For $P=500$~MPa we obtain the changes of the dielectric permittivity of silver $\Delta \epsilon_{\rm{Ag}} = 0.02$ and water $\Delta \epsilon_{\rm{water}} = 0.0035$ (see \cite{Mak16} for numerical details). The stress also leads to the compression of the NA with a decrease in the length and the width $\Delta L \approx 0.5$~nm and $\Delta w \approx 0.05$~nm. Whereas changes in the dielectric permittivity can readily be modelled, simulations of the NA undergoing periodic deformations are challenging when using any technique relying on the discretisation with a mesh. However, as long as linearity holds and Hooke's law remains valid, changes in the volume of the object may be modelled by changing the compressibility of the constituent material \cite{Alu_Khanikaev}. 

In our optical simulations, the role of the compressibility is played by the dielectric permittivity of silver $\epsilon_{\rm{Ag}}$, which is given by the Drude model \cite{Mak16}. The changes in the volume of the NA are taken into account by a factor $\Delta \epsilon_{\rm{vol}}$. In a single CST Microwave Studio simulation run, we obtain the optical spectrum of the NA with a static volume updated using the values of $\Delta L$ and $\Delta w$ and observe a very small but routinely detectable shift in the resonance peak. Then we repeat the simulation for the unperturbed volume NA but vary the value of $\Delta \epsilon_{\rm{vol}}$ to obtain the same spectral shift as in the first simulation. We find $\Delta \epsilon_{\rm{vol}}=-0.005$ that is four times smaller than $\Delta \epsilon_{\rm{Ag}}$ and its sign is opposite to that of $\Delta \epsilon_{\rm{Ag}}$. This implies that the deformation of the NA will decrease the strength of the BLS effect, which is a physical result because a decrease in the volume of the NA leads to a smaller overlap of the near field of the NA with the surrounding medium.

We use a $2$D optical FDTD method (see \cite{Mak16}) to demonstrate an OFC synthesised from the BLS spectrum of NL ultrasound (Fig.~\ref{fig:fig3}). The first six BLS peaks to the left and right of the central (no shift with respect to the frequency of incident light, $405$~THz) peak correspond to the six ultrasound harmonics (Fig.~\ref{fig:fig2}) and they are enhanced by the NA (red solid line in Fig.~\ref{fig:fig3}) with respect to the case of bulk water (green dashed line) used as a reference. Moreover, as established in our case, the cascaded BLS effect leads to the generation of the seventh and so on harmonics seen both in the spectra with and without the NA. Without the NA the intensity of these harmonics quickly drops, but their decrease is significantly less steeper in the presence of the NA. 

In summary, we have demonstrated that harnessing strong NL acoustic interactions in liquids offers new possibilities in the design of ultra-compact sources of OFCs compatible with fibre-optic technology. The spectrum of the obtained OFC is comparable with those produced by microphotonic devices relying on the hyper-parametric oscillation process \cite{Sav16}. The OFCs generated by these devices have a relatively small number of frequency harmonics and a quasi-triangular (in the log scale) power spectrum. However, the number of harmonics obtained with the liquid-state OFC source is higher as compared with the device based on a tapered NA \cite{Mak12}. 

In a NL optical medium, broader and flatter combs may be achieved by mixing two triangular OFCs with different central wavelengths \cite{Tor14}. The same strategy remains applicable when the NL acoustic interactions are exploited. Here, two seed combs may be produced by using ultrasound at different frequencies and mixed in a NL acoustic medium. 

A typical NL-optical microphotonic OFC source has a cubic centimetre volume and operates at $\sim 20$~mW of laser power \cite{Hau14}, but the total consumed power is $\sim 1$~W \cite{Sav16}. Because NL optical interactions at the nanoscale are weaker than at the microscale, the power needs to be increased when plasmonic NAs are employed. We estimate that the OFC source based on a single tapered NA requires at least $3$~W of continuous wave laser power, which is a challenging requirement.

The proposed ultrasound-based OFC source does not need high laser power, but it demands ultrasound pressures of $\sim 0.5$~GPa. When the ultrasound focal area dimensions are only limited by the diffraction (half the wavelength) limit, we estimate that for the synthesis of the comb shown in Fig.~\ref{fig:fig3} one needs the acoustic power of $\sim 10$~W at $f_{\rm{us}}= 100$~MHz. 

Acoustic power of up to $\sim 1$~kW is readily achievable with medical high-intensity focused ultrasound (HIFU) transducers that have a $\sim 50\%$ electrical-to-acoustic conversion efficiency \cite{hifu}. Moreover, it has been shown that the focal area of the transducer may be reduced by using acoustic metamaterials, which allows focusing ultrasound into deep-subwavelength, $1/40$ of the incident wavelength spots \cite{Maznev}. Thus, the required acoustic power may be decreased to $\sim 20$~mW.

High pressure ultrasound may also be produced by strong pulses emitted by collapsing gas bubbles in water \cite{bubble}. The application of bubbles is attractive also because the nonlinear parameter $\beta$ of water with air bubbles is three orders of magnitude larger than that of without bubbles \cite{Gurbatov}. Such a high acoustic nonlinearity will result in stronger NL acoustic interactions and lower acoustic power required for the operation of the device.

This work was supported by Australian Research Council (ARC) through its Centre of Excellence for Nanoscale BioPhotonics (CE140100003). This research was undertaken on the NCI National Facility in Canberra, Australia, which is supported by the Australian Commonwealth Government.


\end{document}